**Compact optical polarization-insensitive zoom metalens-doublet**

*Yunxuan Wei, Yuxi Wang, Xing Feng, Shunyuan Xiao, Zhaokun Wang, Tie Hu, Manchen Hu, Martin Wegener, Ming Zhao\*, Jinsong Xia\*, and Zhenyu Yang\**

Y. X. Wei, Dr. Y. X. Wang, Dr. X. Feng, S. Y. Xiao, Dr. Z. K. Wang, T. Hu, M. C. Hu, Prof. M. Zhao, Prof. J. S. Xia and Prof. Z. Y. Yang
Wuhan National Laboratory for Optoelectronics, School of Optical and Electronic Information, Huazhong University of Science and Technology (HUST), Wuhan 430074, China
E-mail: zyang@hust.edu.cn, jsxia@hust.edu.cn, zhaoming@hust.edu.cn

Prof. M. Wegener
Institute of Nanotechnology and Institute of Applied Physics, Karlsruhe Institute of Technology (KIT), Karlsruhe 76128, Germany

Metasurface-based lenses (metalenses) offer specific conceptual advantages compared to ordinary refractive lenses. For example, it is possible to tune the focal length of a metalens doublet by varying the relative angle between the two metalenses while fixing their distance, leading to an extremely compact zoom lens. An improved polarization-insensitive design based on silicon-nanocylinders on silica substrates is presented. This design is realized and characterized experimentally at 1550 nm wavelength. By varying the relative angle between the metalenses in steps of 10 degrees, tuning of the doublet focal length is demonstrated from $-54$ mm to $\pm 3$ mm to $+54$ mm. This results in a zoom factor of an imaging system varying between 1 and 18. For positive focal lengths, the doublet focusing efficiency has a minimum of 34% and a maximum of 83%. Experiment and theory are in very good agreement.

## 1. Introduction

Optical zoom lenses have wide applications in cameras, mobile phones, augmented reality, and microscopy. Most traditional zoom systems rely on varying the axial distance between two or more ordinary refractive lenses to change the effective focal length,[1] making such systems complex and large. Many alternatives have been discussed. This includes Alvarez lens systems based on lateral displacements of two refractive free-form surfaces,[2,3] liquid-crystal based devices,[4-9] and deformable lenses based on liquids[10-13] or other elastic constituents.[14]

Tunable metasurfaces,[15-19] with the ability to control the wave front dynamically under a subwavelength scale, enables ultrathin adjustable devices. Tunable metasurface-based lenses or "metalenses"[20-28] have been discussed as well. Examples are metalenses on elastic substrates,[29-31] Alvarez metalens systems,[32] and tuning of the metalens focal length by stimuli such as temperature[33,34] or static electric fields combined with dielectric elastomer actuators[35] or microelectromechanical systems.[36,37]

More recently, the concept of rotatory metalens zoom doublets has been introduced.[38-41] Here, the focal length can be varied by changing the relative angle between the two metasurfaces while fixing their distance. This compact, simple, and robust design – which has no counterpart for ordinary refractive lenses – has been discussed theoretically.[38-40] Early experiments at microwave frequencies have demonstrated the validity of the concept, albeit in a polarization-dependent manner.[41] The novelty of the present paper lies in presenting an improved polarization-insensitive theoretical design, and in providing corresponding experimental focusing characterizations as well as imaging demonstrations at telecommunication wavelengths, with a zoom varying by as much as a factor of 18.

## 2. Designs and Principles

The operation principle of the metalens doublet is illustrated in **Figure 1**a. In order to establish the relationship between focal length and relative angle between the two metasurfaces, we construct an output wave front from the doublet that depends on the angle. We choose the phase profile of a spherical defocusing lens, which results in smaller phase errors than that of a focusing lens (see Section S1). For the two metasurfaces in the doublet, we choose the two phase profiles

$$\varphi_1(r,\theta_0) = round[\frac{1}{\lambda F_0}r^2](\theta_0 + C(r)) \tag{1}$$

and

$$\varphi_2(r,\theta_0) = -round\left[\frac{1}{\lambda F_0}r^2\right](\theta_0 + C(r)). \tag{2}$$

Here, $r$ is the radial coordinate, $\lambda$ is the operating wavelength in free space, $F_0$ is a reference focal length, and $\theta_0$ is a reference rotation angle. The function $round[...]$ replaces its argument by the nearest integer. Thereby, this function ensures that the phase shift varies by an integer multiple of $2\pi$ when increasing $\theta_0$ from 0 to $2\pi$. Deviations from this behavior would result in discontinuities around $\theta_0 = 2\pi$, causing unwanted phase errors of the doublet. The function $C(r)$ is zero for all $r$ for the ideal case of zero axial distance between the two metasurfaces. A non-zero choice of this function serves to compensate distortions of the focal spot that arise from the finite distance. We have optimized this function to $C(r) = \frac{\pi}{12}round(\frac{r^2}{\lambda F_0})$. A comparison to the case of $C(r) = 0$ is given in Section S2.

Combining the two metasurfaces and neglecting the finite distance between them, the phase profile of the doublet is given by

$$\varphi(r,\theta) = \varphi_1(r,\theta_0) + \varphi_2(r,\theta_0 + \theta) = -round\left[\frac{1}{\lambda F_0}r^2\right]\theta. \tag{3}$$

For reference, the phase profile of a spherical lens with focal length $F$ is given by $\varphi_0(r) = -\left(\frac{1}{\lambda F}r^2\right)\pi$. Comparing $\varphi(r,\theta)$ and $\varphi_0(r)$, we find that $\varphi(r,\theta)$, which is shown in Figure 1b, is a quantized version of $\varphi_0(r)$. Moreover, the focal length of the doublet becomes a function of the rotation angle $\theta$ according to

$$F(\theta) = \frac{\pi}{\theta}F_0. \tag{4}$$

We define the ratio $k = \frac{\pi}{\theta}$ as the zoom power of the doublet. The mentioned quantization leads to a decrease of the diffraction efficiency. Following the theory of multilevel diffractive lenses,[42] the diffraction efficiency is given by $\eta = (\text{sinc}(1/n))^2$, where $n = 2\pi/\theta$ is the number of quantized levels. For example, we obtain $\eta = 40.5\%$ and $k = 1$ for $\theta = \pi$, $\eta = 81.1\%$ and $k = 2$ for $\theta = \pi/2$, and $\eta = 95.0\%$ and $k = 4$ for $\theta = \pi/4$. Clearly, the diffraction efficiency increases with increasing $F \propto 1/\theta$.

We target a free-space operation wavelength of the metalens zoom doublet of $\lambda = 1550$ nm. To achieve the above phase profiles in this wavelength regime, we consider metasurfaces composed of silicon (Si) nanocylinders, with cylinder radius $R$ and height $H$, arranged on a square lattice with period $P$ on a silica (SiO$_2$) substrate as shown in **Figure 2**a. The cylindrical geometry of the Si nanopillars combined with the square array ensures insensitivity with respect to the polarization of the incident light. The material combination Si and SiO$_2$ provides high refractive-index contrast and high optical transmission at the same time. The resulting optical response has been calculated and optimized by using a finite-difference time-domain (FDTD) approach.

Because the two metasurfaces in the doublet are placed face to face, light impinges from the bottom of the first metasurface and from the top of the second metasurface. As diffraction of light is involved, the corresponding optical transmissions are not necessarily identical. Therefore, we investigate both propagation directions. Our optimization of the parameters $R$, $H$, and $P$ follows two rules. First, we optimize the three parameters to obtain high transmittance for both propagation directions and full phase control. Second, the response of the nanocylinders should depend only weakly on the angle of incidence, as oblique incidence plays a role in most practical circumstances. We arrive at $P = 600$ nm and $H = 700$ nm at $\lambda = 1550$ nm.

The transmittances and phase shifts of the nanocylinders versus $R$ under normal incidence of light are shown in Figure 2a. Responses under different incident angles are depicted in Figure 2b and c. With the radius $R$ varying from 60 nm to 260 nm, a phase shift of nearly $2\pi$ and a transmittance above 75% is achieved for normal incidence. Furthermore, within the range if radii investigated by us, most Si nanocylinders exhibit only a weak dependence on the incidence angle within an angular range of 20°.

**3. Results and Discussions**

**3.1. Focusing with the metalens doublet**

To experimentally test the above design, we have fabricated a corresponding metasurface doublet with a diameter of 1 mm and a minimum focal length of 3 mm. Details of the fabrication and the designed phase profile are discussed and shown in Section S3 and Figure S4. Optical micrographs of two fabricated metasurfaces and scanning electron micrographs for one of them are shown in Figure 2d and Figure 2e,f, respectively. A SiO$_2$ layer with a thickness of about 2.5 µm has additionally been deposited around the nanocylinder area for both metasurfaces to protect the nanocylinders from being scratched, which corresponds to the yellow ring in Figure 1a. Additional alignment markers have been applied to guide the rotation alignment.

The focusing behavior of the doublet was measured under collimated illumination at a wavelength of $\lambda = 1550$ nm. Details of the experimental characterization setup are given in Figure S5 and S6.

**Figure 3**a depicts the measured (normalized) intensity of light, depicted in the $xz$ plane and in the focal plane under six different rotation angles $\theta$. The focal length changes from 3 mm to 54 mm. Correspondingly, the doublet zooming power varies from $k = 1$ to $k = 18$, and the doublet numerical aperture (NA) from 0.164 to 0.001. At each value of $\theta$, the position of the focal plane fits well with the designed focal length. As to be expected, the depth and the size of the focus increases with increasing focal length. We have also measured the focal length of the doublet for varying the angle $\theta$ in steps of 10 degrees. The results shown in Figure 3b cover both positive and negative focal lengths. The average relative deviation between experiment and theory is as small as 1.7%. This error can be traced back to errors in experimentally determining $\theta$.

Figure 3c characterizes the focusing efficiency and the full width at half maximum (FWHM) for positive focal lengths. Such measurements are not easily possible for negative focal lengths. The focusing efficiency is defined as the power gathered in a circular area with a diameter of 2×FWHM and the total power in a circular area with 1 mm diameter in the focal plane. The center of both circular areas is the point of the maximum intensity. The average measured FWHM is 27% larger than the diffraction limit of $0.514\ \lambda/\mathrm{NA}$. For $\theta = 10°$, the FWHM is 60% larger than the diffraction limit. This behavior results from a weak focusing behavior at a small numerical aperture of $\mathrm{NA} = 0.001$. The focusing efficiency exhibits a minimum of $\eta = 23\%$ at $\theta = 180°$, where the focal length has its minimum of 3 mm, and increases to $\eta = 82\%$ at $\theta = 10°$, where the focal length has its maximum of 54 mm. This overall behavior is expected from the doublet phase profile discussed above. The average focusing efficiency is 54%, which is slightly smaller than the numerically determined value of 63%. We attribute this difference partly to the finite axial separation of the two metasurfaces, which has been neglected in the design process and which leads to phase distortions. Fabrication imperfections of the Si nanocylinders and their arrangement may also contribute. In addition, the remaining dependence on the angle of incidence (cf. Figure 2b,c), which has not been accounted for in the design process, is likely yet another contribution.

### 3.2. Imaging with the metalens doublet

We have also experimentally characterized the zooming ability of the metalens doublet in imaging experiments using a $4f$ system, which consists of the metalens doublet and a conventional lens with a focal length of 200 mm (see Figure S7 for details). We have manufactured a dedicated test sample (see Figure S8). The narrowest line width of the ideal image of this sample is about $1.1\times$ the focal spot FWHM for $\theta = 180°$. The slenderest part of the "HUST" logo is still close to the diffraction limit, making it dimmer than other areas. The zooming image was recorded under the rotation angles $\theta$ of 10°, 20°, 30°, 50°, 90°, and 180°, corresponding to zooming powers of $k = 18, 9, 6, 3.6, 2$, and 1. In order to avoid light going around the doublet aperture contributing to the images, which would reduce image contrast, the whole sample image has been captured part by part with an aperture and then reconstructed. Reconstructed images are depicted in **Figure 4**a-f. Under these conditions, the overall resolution of the system varies only slightly. We have also magnified the images for $\theta = 180°$, $90°$, and $50°$ to the same ideal magnification of $k = 9$ in Figure 4g-h. All of them are clearly imaged with approximately the same sizes. The test sample is clearly imaged for all zoom factors, spanning a range from $k = +1$ to $k = +18$.

## 4. Conclusion

In summary, we have presented experimental results following an optimized design of a zoom metalens doublet, based on rotating two metasurfaces with respect to each other at fixed distance between them, at an operation wavelength of 1550 nm. The focal length could be adjusted from $\pm 3$ mm to $\pm 54$ mm, corresponding to extremal zoom factors of $\pm 18 \times$ at an average focusing efficiency of 54%. Such compact and easy-to-use tunable zoom metalens doublets can be scaled to other operation wavelengths and could find niche applications where compactness is key and only a narrow wavelength range is needed.

**Acknowledgements**

This work was funded by the National Natural Science Foundation of China (NSFC, grant no. 61835008 and grant no. 11574102), the Fundamental Research Funds for the Central Universities (grant no. 2019kfyXKJC038), and the State Key Laboratory of Advanced Optical Communication Systems and Networks project, Shanghai Jiao Tong University, China (grant no. 2019GZKF03001). We are grateful for device fabrication support from the Center of Micro-Fabrication and Characterization (CMFC) in the Wuhan National Laboratory for Optoelectronics (WNLO) of Huazhong University of Science and Technology (HUST). M.W. acknowledges support by the Deutsche Forschungsgemeinschaft (DFG, German Research Foundation) under Germany's Excellence Strategy via the Excellence Cluster "3D Matter Made to Order" (EXC-2082/1-390761711), which has also been supported by the Carl Zeiss Foundation through the "Carl-Zeiss-Focus@HEiKA".

Yunxuan Wei, Yuxi Wang and Xing Feng contributed equally to this work.

**Author contributions**

Y.X.Wei had the original idea, conceived the study, finished the simulations; Y.X.Wang fabricated the samples and analyzed them; X.Feng performed the measurements with the help from Y.X.Wei, S.Y.Xiao, Z.K.Wang and T.Hu; J.S.Xia supervised the fabrication; M.Zhao, Z.Y.Yang, and M.Wegener supervised the work and the manuscript writing. All authors discussed the results. Y.X.Wei wrote a first draft of the manuscript, which was refined by contributions from all authors.

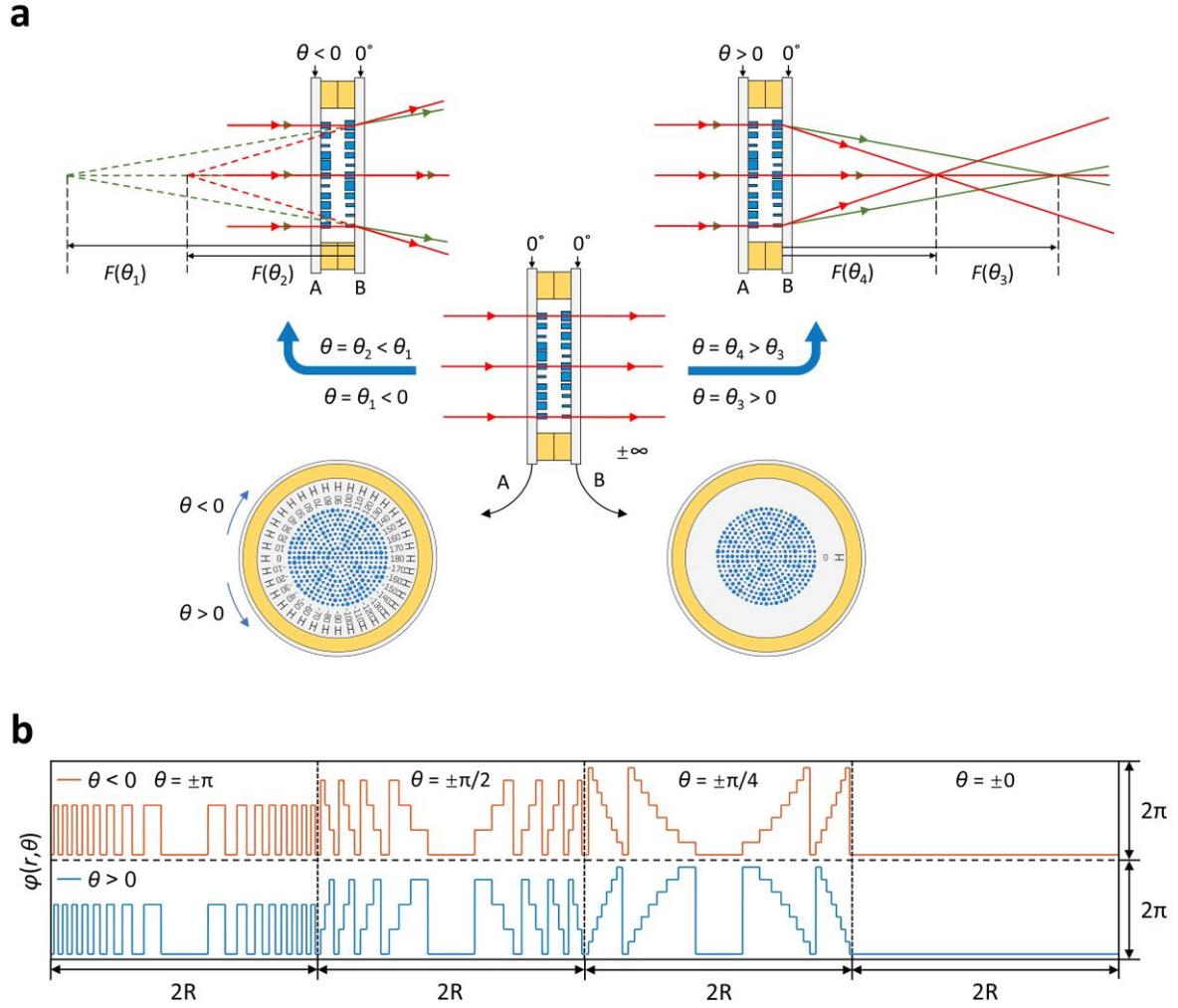

**Figure 1.** Illustration of the metalens doublet and designed wavefront output. a) Scheme of the zooming imaging doublet consisting of two metasurfaces. The focal length of the doublet changes continuously when varying the relative angle $\theta$. For $\theta > 0$, the doublet works as a positive lens, for $\theta < 0$ as a negative lens. The "H" is merely an alignment marker. b) Output phase profile of the doublet for different values of $\theta$. The quantization of the phase changes when varying $\theta$.

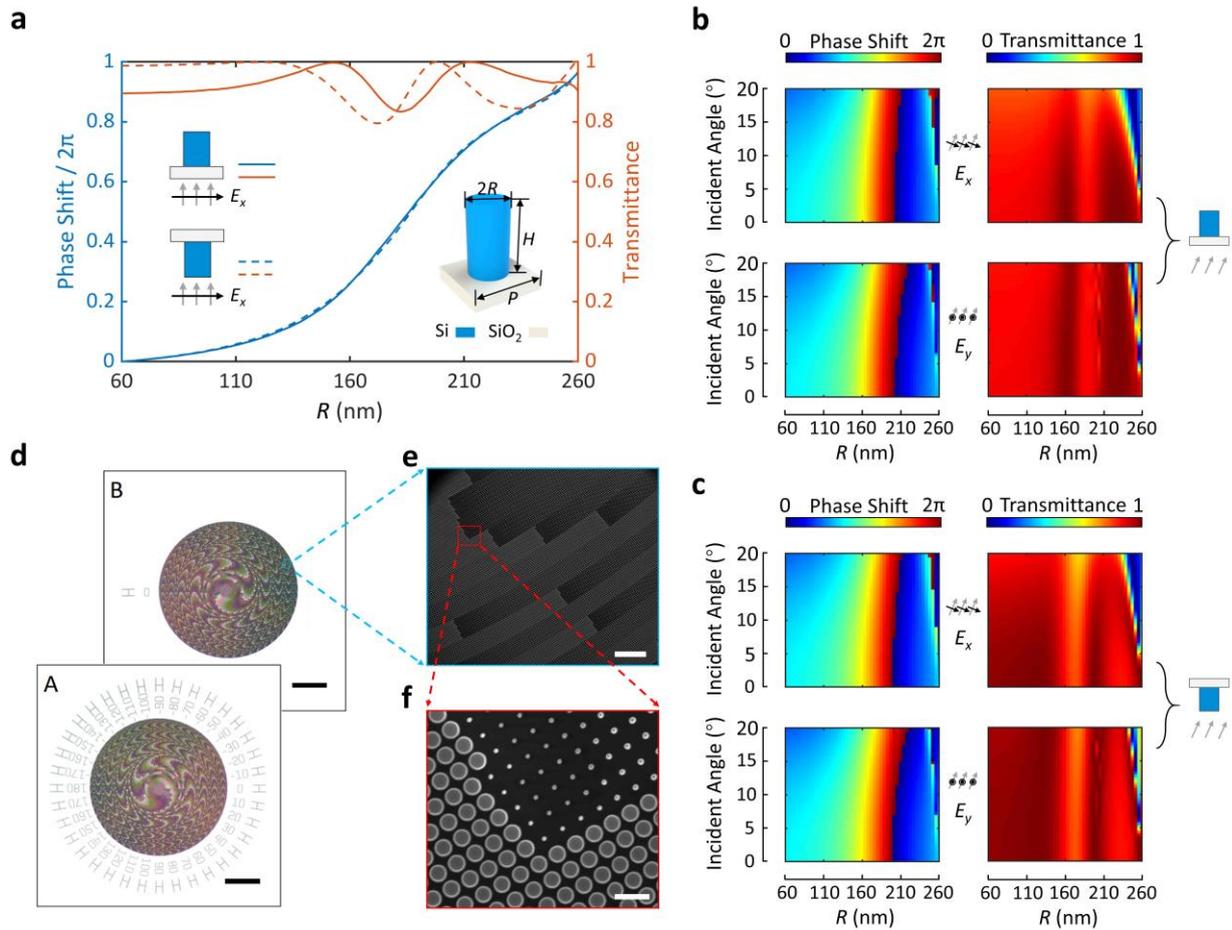

**Figure 2.** Design of the silicon nanocylinders. a) Calculated phase shift and transmittance of a square array of nanocylinders with varying radius $R$, fixed height $H = 700$ nm, fixed lattice constant $P = 600$ nm, for linearly polarized incident light, and for normal incidence with respect to the silica substrate plane. Results are shown for light impinging from the bottom (solid curve) and the top (dashed curve) with respect to the substrate. b) Calculated angular dependence of phase shift and transmittance with light impinging from the bottom. The upper (lower) two panels correspond to linear incident polarization oriented along the $x$-direction ($y$-direction). c) Dependence of phase shift and transmittance versus angle of incidence for light impinging from the top, again for the two different orthogonal linear polarizations of light. d) Optical micrograph of the fabricated metalens doublet. Scale bar is 250 µm. e) Scanning electron micrograph of the area marked in (d). Scale bar is 7.5 µm. f) Scanning electron micrograph of the area marked in (e). Scale bar is 1 µm.

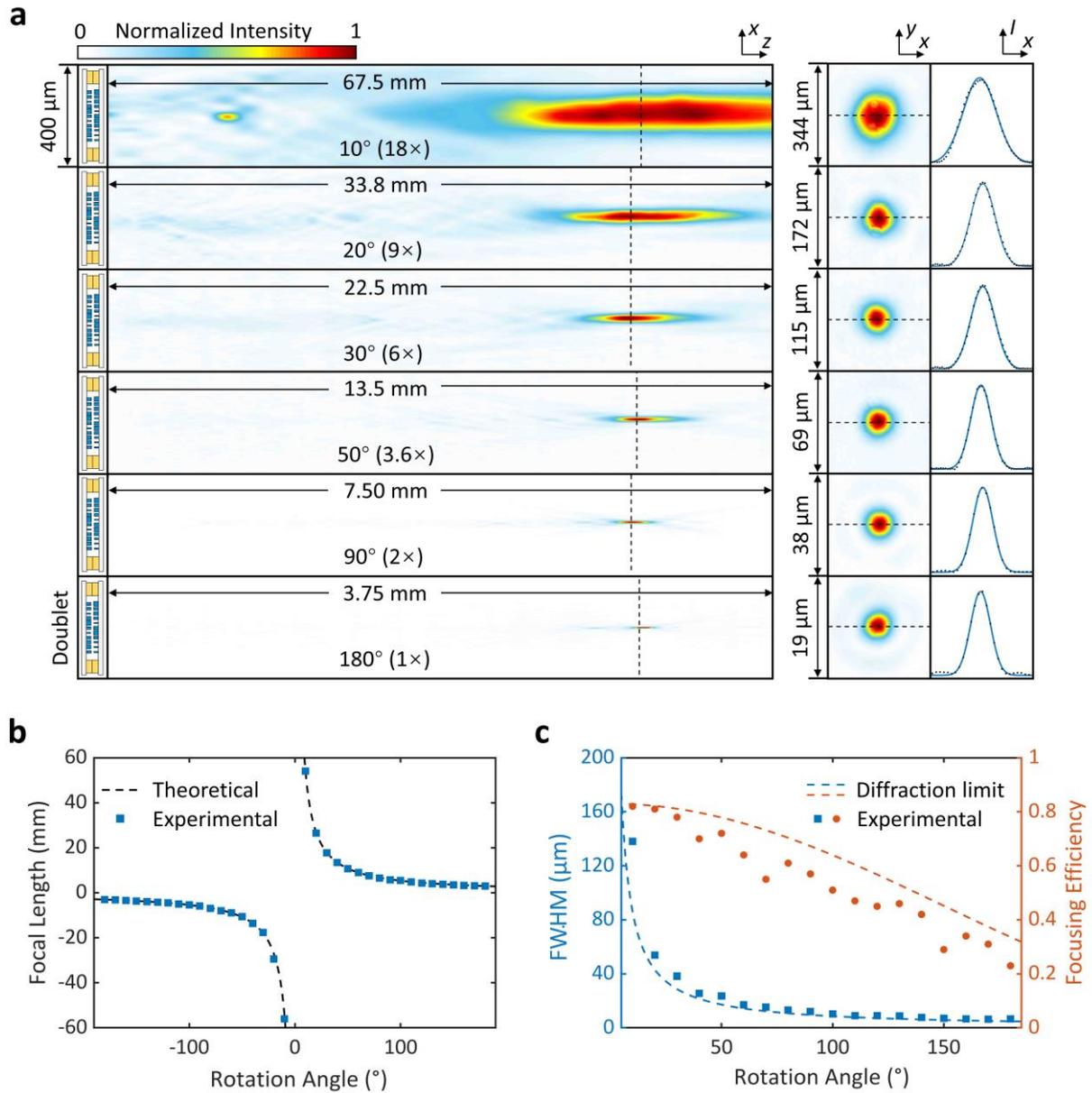

**Figure 3.** Measured focusing behaviors and corresponding key properties of the metalens doublet. a) Normalized intensity distributions for six different rotation angles $\theta$. The zoom changes by a factor of 18. The focal planes are highlighted by the dashed black lines. Cuts in these planes are shown on the right-hand side. All of the pictures are with appropriately changed scales marked on them. b) Theoretical (dashed curve) and experimental (blue symbols) focal length versus the rtation angle $\theta$. c) Determined full width at half maximum (FWHM) and focusing efficiency $\eta$ versus $\theta$ for positive focal lengths. Theory (dashed curves) and experiment (symbols) agree well.

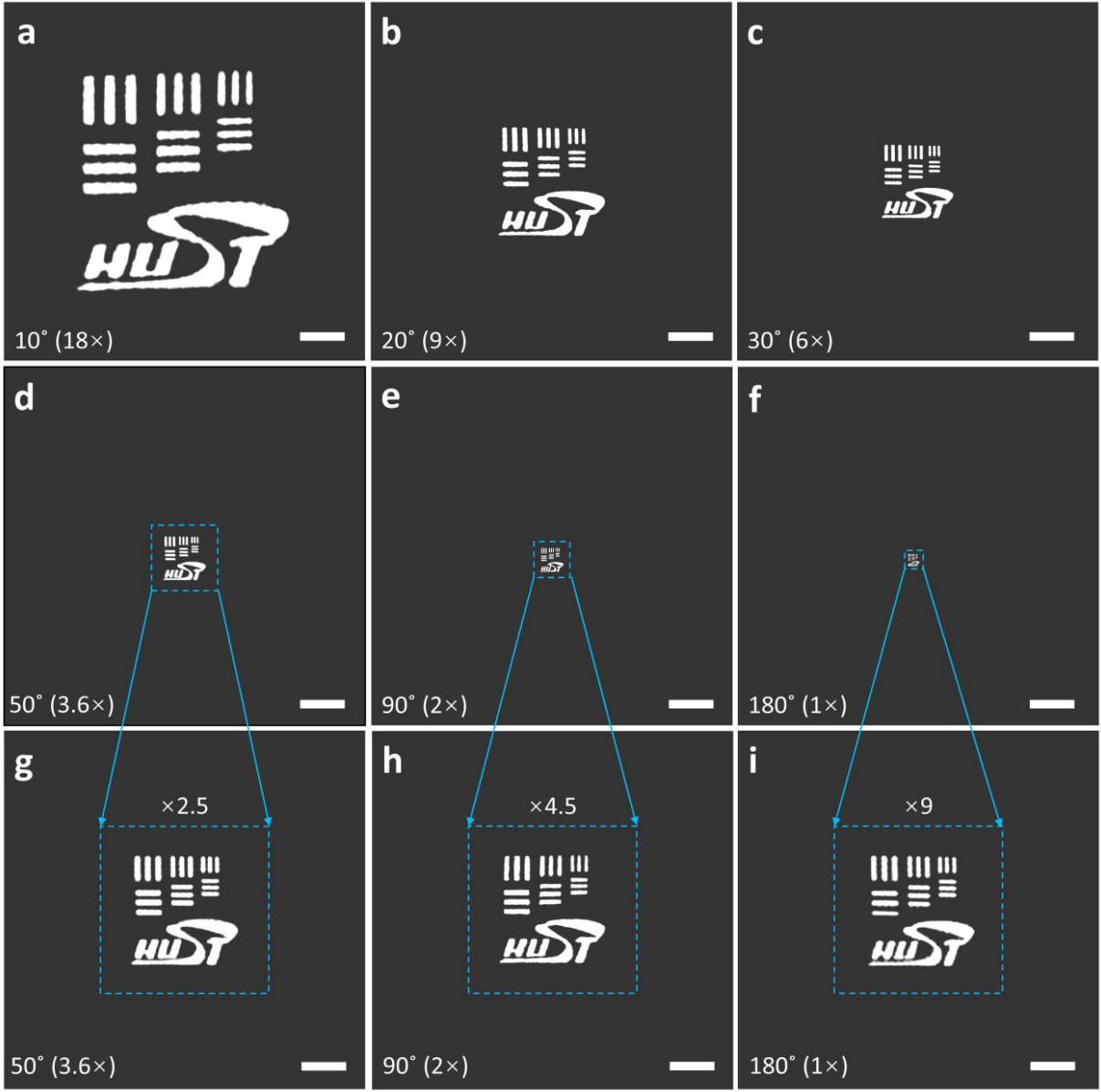

**Figure 4.** Zooming images measured with a 4*f* system consisting of the metalens doublet and a conventional refractive lens. a-f) Different rotations angles $\theta$ as indicated in the lower left-hand side corners. Magnification factor in parentheses. The total zoom range is 18-fold. Scale bar is 1 mm. g-i) Magnified images for $\theta = 50°$, 90°, and 180°. The magnification factor is chosen as $9/k$, with the zoom factor $k(\theta)$. Scale bars are 400 μm in (g), 222 μm in (h), and 111 μm in (i).

# Supporting Information

**Compact optical polarization-insensitive zoom metalens-doublet**

*Yunxuan Wei, Yuxi Wang, Xing Feng, Shunyuan Xiao, Zhaokun Wang, Tie Hu, Manchen Hu, Martin Wegener, Ming Zhao\*, Jinsong Xia\*, and Zhenyu Yang\**

**Section S1.** Comparison of the phase error for different wave fronts

In our design, the phase profile of the doublet is chosen as the wave front of a spherical defocusing lens. Here, we explain about this with a comparison of the phase error of two kinds of designs.

Despite the impact of the quantization, when designed with the phase profile of a spherical lens $\varphi_0(r) = -\frac{\pi}{\lambda F_0}r^2$ in the main text, the output wave front of the metalens doublet is

$$\varphi_s(r,\theta) = -\frac{\theta}{\lambda F_0}r^2 = -\frac{\pi F_0}{\lambda}\frac{F_0}{F(\theta)}\left(\frac{r}{F_0}\right)^2. \tag{1}$$

Meantime, when designed with the phase profile of a focusing lens $\varphi_{0'}(r) = -\frac{2\pi}{\lambda}\left(\sqrt{r^2 + F_0^2} - F_0\right)$, the phase profiles of two metasurfaces are $\varphi_{1'}(r,\theta_0) = -round[\varphi_{0'}(r)/\pi]\theta_0$ and $\varphi_{2'}(r,\theta_0) = round[\varphi_{0'}(r)/\pi]\theta_0$. Despite of the impact of the quantization, the output wave front of the metalens doublet is

$$\varphi_f(r,\theta) = \frac{\varphi_{0'}(r)}{\pi}\theta = -\frac{2\pi F_0}{\lambda}\frac{F_0}{F(\theta)}\left(\sqrt{\left(\frac{r}{F_0}\right)^2 + 1} - 1\right). \tag{2}$$

However, under different focal length $F(\theta)$, the phase profile of a focusing lens is

$$\varphi_a(r,\theta) = -\frac{2\pi}{\lambda}\left(\sqrt{r^2 + F(\theta)^2} - F(\theta)\right) = -\frac{2\pi F_0}{\lambda}\left(\sqrt{\left(\frac{r}{F_0}\right)^2 + \left(\frac{F(\theta)}{F_0}\right)^2} - \frac{F(\theta)}{F_0}\right). \quad (3)$$

Here, we regard $\varphi_a(r,\theta)$ as an accurate focusing wave front. Then the ratio of phase error can be defined as $\mu_s = |\frac{\varphi_s - \varphi_a}{\varphi_a}|$ for $\varphi_s(r,\theta)$ and $\mu_f = |\frac{\varphi_f - \varphi_a}{\varphi_a}|$ for $\varphi_f(r,\theta)$. The calculation results are plotted in Figure S1. It is obvious that $\varphi_f(r,\theta)$ introduces more phase errors for the doublet, thus $\varphi_s(r,\theta)$ which results from the design with the spherical defocusing lens can be a better choice.

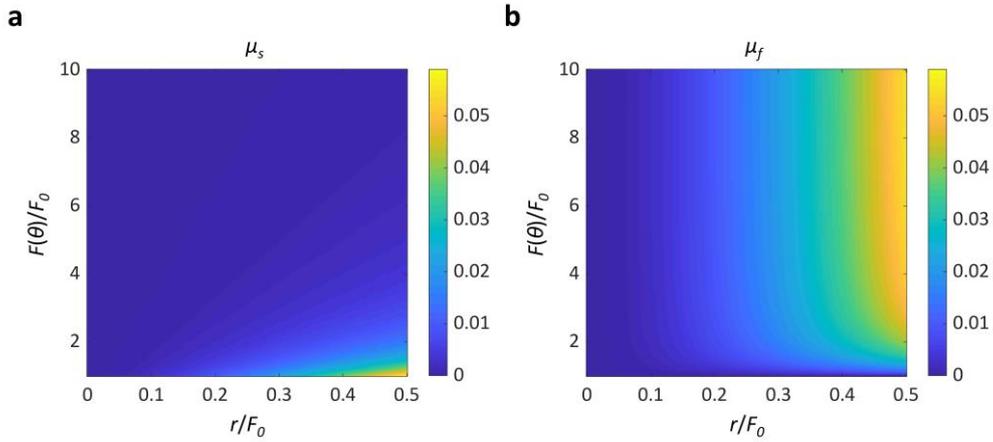

**Figure S1.** Ratios of phase errors for the doublet based on (a) spherical defocusing lens and (b) focusing lens.

**Section S2.** Design and experimental results of a metalens doublet with $C(r) = 0$

Designed phase profiles and optical micrographs of the doublet is shown in Figure S2a-d. We have measured the focal spot of the metalens doublet under the two rotation angles of 180° and 90° with the same experimental setup as described in the main text. The experimental results are shown in Figure S2e-h. The results of the doublet in the main text are repeated in Figure S2i-l for convenient comparison. For the doublet with $C(r) = 0$, the shapes of both the focal spot and its side lobe are distorted with asymmetrical characteristics. Such distortions will negatively impact the performance of the doublet.

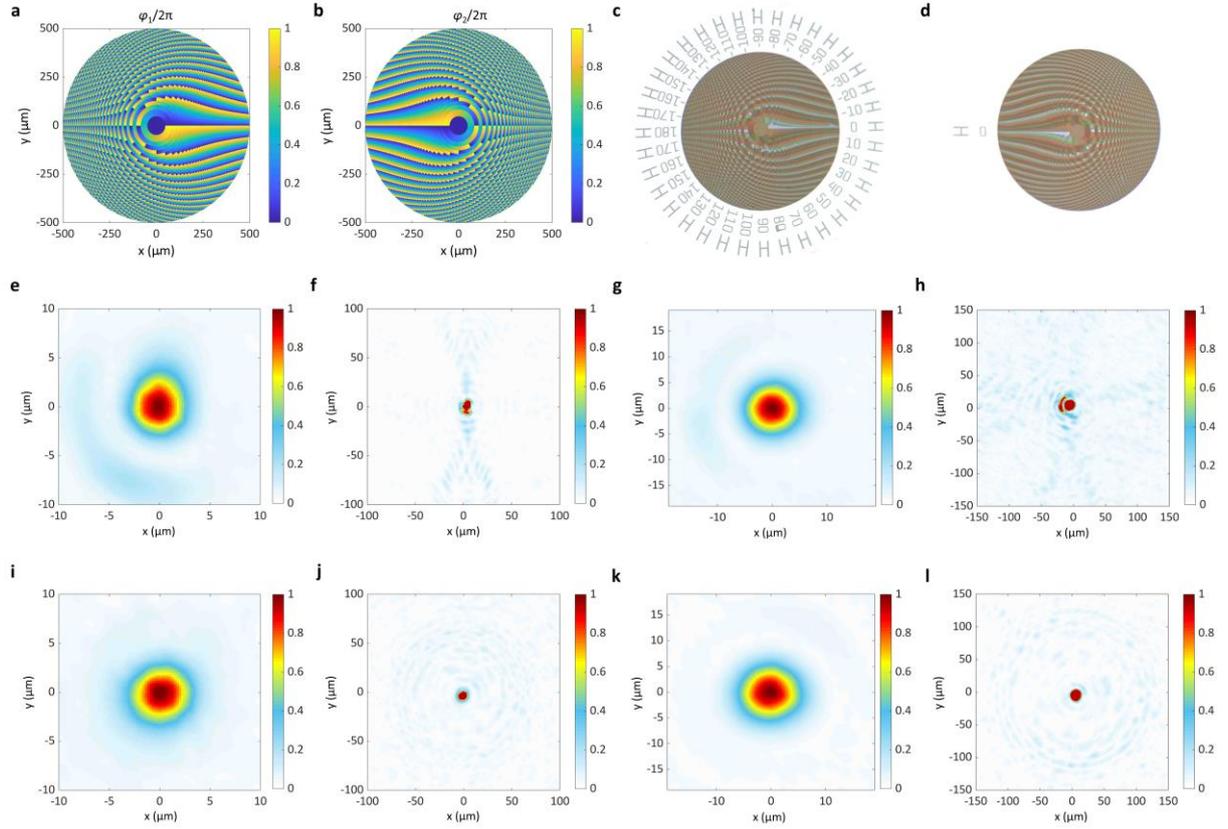

**Figure S2**. Design and experimental results of a metalens doublet with $C(r) = 0$. a,b) Designed phase profiles of two individual metasurfaces. c,d) Optical micrographs of two fabricated metasurfaces. e-h) Measured focal spots of the doublet designed with $C(r) = 0$ under the rotation angle of (e, f) 180° and (g, h) 90°. i-j) Measured focal spot of the doublet in the main text. The rotation angle is (i, j) 180° and (k, l) 90°. The intensity of the focal spot is saturated in panels (f), (j), (l), (h) in order to make the side lobes better visible.

**Section S3.** Details of the fabrication of the metalens doublet

The fabrication of the doublet starts from a double-polished silicon-on-insulator (SOI) wafer with a 700 nm top silicon (Si) layer and 2 μm buried oxide layer. Key procedures are illustrated in the figure below. First, a chromium (Cr) layer with a thickness of 20 nm is deposited by electron-beam evaporation (EBE, Ohmiker-50B) on top of the Si layer as a hard mask. Next, a 200 nm photoresist layer (CSAR62) is spin-coated onto the top of the Cr layer. The pattern of the tunable metalens is written by electron-beam lithography (EBL, Vistec: EBPG 5000 Plus) into the photoresist layer. After development, the pattern is then transferred into the Cr hard mask layer by inductively coupled plasma etching (ICP, Oxford Plasmalab: System100-ICP-180), and the residual photoresist is stripped off by an oxygen plasma stripper (Diener electronic: PICO plasma stripper). Finally, the pattern is transferred into the Si layer by the next ICP process, and the remaining Cr is removed. The Cr layer is utilized as a hard mask because of the extremely high etching selectivity between Cr and Si.

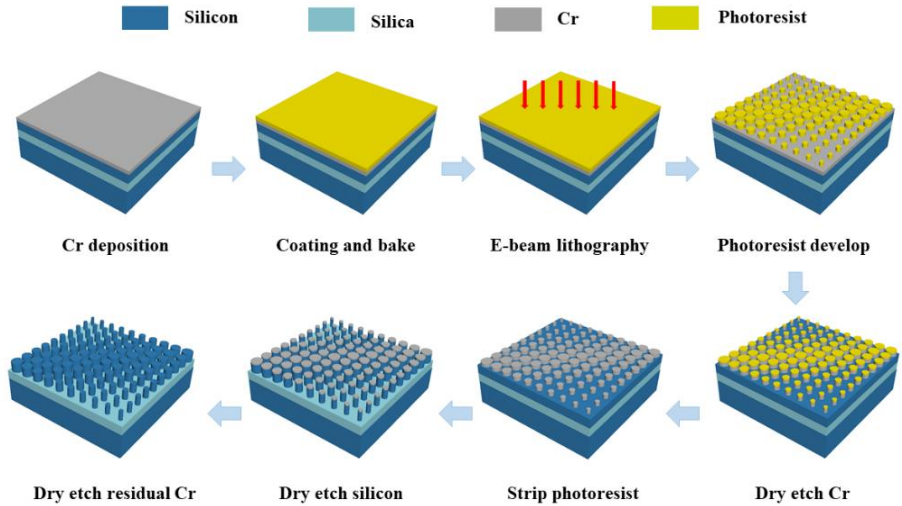

**Figure S3.** Fabrication flow for the metalens doublet.

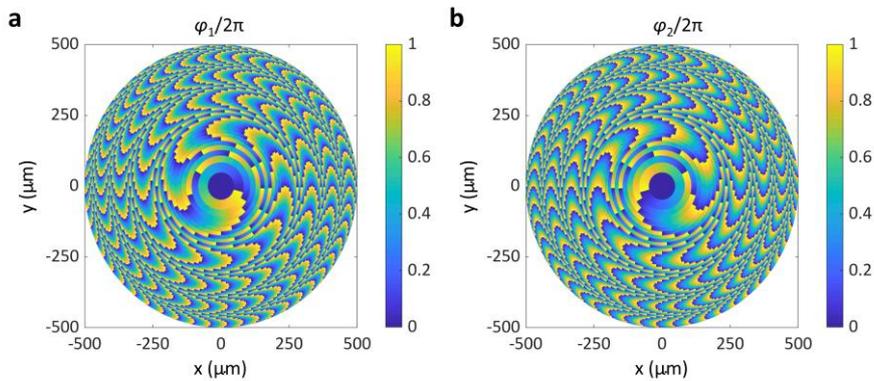

**Figure S4.** Phase profiles of the fabricated doublet. The phase profile of the first and second metasurface are shown in (a) and (b).

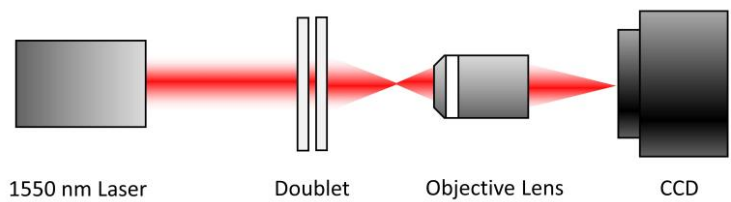

**Figure S5.** Optical setup for measurements for positive focal length. The incident beam is derived from a continuous-wave laser (JOINWIT: JW8002) with a free-space wavelength of 1550 nm. The incident light is focused by the doublet metalens and imaged onto the InGaAs-

based camera (Xenics: Bobcat-320-Gige) by a microscope objective. To determine the focal length, we measure the light intensity on the optical axis by moving the metalens doublet along the optical axis. In order to perform these measurements for rather different spot sizes, we have chosen different microscope objective lenses, namely $4\times$, NA $= 0.1$ and $10\times$, NA $= 0.25$.

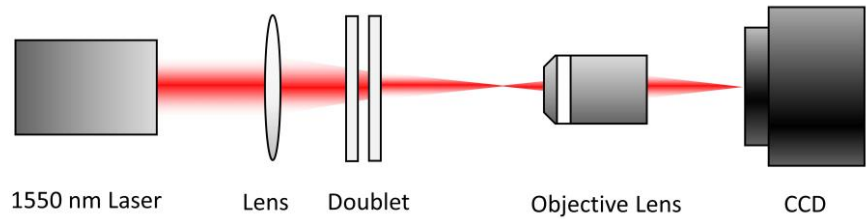

**Figure S6.** Optical setup for measurements for negative focal length. As the negative focal length is difficult to measure directly, a conventional lens (with 40 mm focal length) is inserted 37.27 mm before the doublet metalens. Together, they effectively form a positive optical lens. By measuring its effective focal length, the (negative) focal length of the metalens doublet can be calculated.

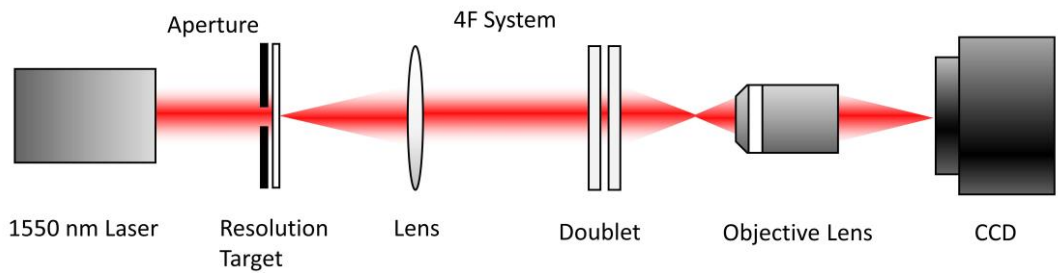

**Figure S7.** Optical imaging setup. A $4f$ system is formed by the metalens doublet and a conventional lens (200 mm focal length). The aperture in front of the resolution target is used to improve the image quality. When one part of the resolution target is captured, the target will be moved and the next part is aligned to the aperture for capturing. With changing focal length of the metalens doublet, the image size on the CCD camera would vary considerably. To compensate for that effect, we have chosen different magnifications of the objective lens, namely $4\times$, $10\times$, and $20\times$.

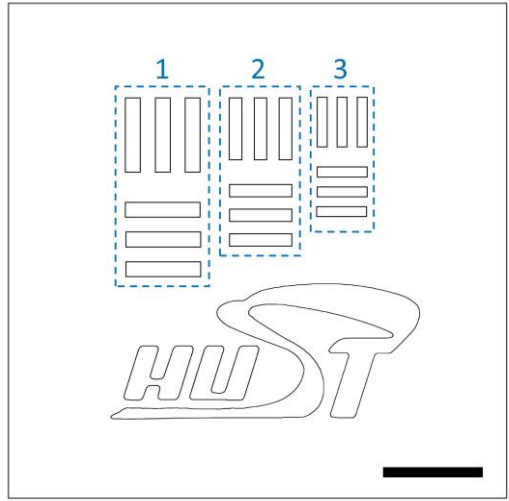

**Figure S8.** Details of the designed resolution target. For line groups 1, 2 and 3, the line width are 0.75 mm, 0.625 mm and 0.5 mm. The line lengths are 3.75 mm, 3.125 mm, and 2.5 mm. The periods are 1.5 mm, 1.25 mm, and 1 mm. The target was fabricated on a thick black cardboard by using laser marking. Scale bar is 5 mm.